\documentstyle[preprint,tighten,aps]{revtex}
\begin{document}

\title{Quantum Cryptographic Network based on Quantum Memories}   

\author{Eli Biham}
\address{Computer Science Department\\ Technion\\ Haifa
32000, Israel} 
\author{Bruno Huttner}
\address{Group of Applied Physics\\University of Geneva\\
CH-1211, Geneva 4, Switzerland} 
\author{Tal Mor}
\address{Department of Physics\\ Technion\\
Haifa 32000, Israel} 
\date{3.Mar.1996}

\maketitle

\begin{abstract}

Quantum correlations between two particles 
show non-classical properties 
which can be used for providing secure transmission 
of information.
We present a quantum cryptographic system, 
in which users store particles in quantum memories
kept in a transmission center. 
Correlations between the particles stored by two users 
are created upon 
request by projecting their product state  
onto a fully entangled state. 
Our system allows for secure communication between
any pair of users who have particles in the same center.
Unlike other quantum cryptographic systems, it can work 
without quantum channels and is suitable for  
building a quantum cryptographic network.  
We also present a modified system with many centers.
\end{abstract}


\section{Introduction and motivation}

The main goal of cryptography, the secure transmission of messages,
can be achieved using a secret key known only to the sender, Alice,
and the receiver, Bob.
The only known way which might allow two users 
to create an unconditionally secret key without sharing 
any common information in advance 
is quantum cryptography  
\cite{BB84,Eke91,BBM,Ben92,BBBSS,BW}. 
In quantum cryptographic schemes Alice 
uses non-orthogonal quantum states (transmitted
through a quantum channel) to transfer
the key to Bob. Such states cannot be cloned hence any 
attempt by an eavesdropper, known as Eve, to get information on the
key disturbs the transmitted signals and induces 
noise.
This noise will be detected by Alice and Bob during the second stage
of the transmission, which includes discussion over a public channel. 
The alternative to quantum key distribution schemes, Public
Key Cryptography~\cite{DH,RSA}, relies 
on computational complexity assumptions such as the 
difficulty of factoring. To date,
none of the existing public key cryptosystems  
is proven secure, even against 
attacker with limited 
computation power.
Moreover, it was recently shown~\cite{Shor} that these complexity
assumptions may not hold for a quantum computer (for example, a
quantum computer should enable fast factorization). 
This implies that many public 
key cryptosystems,  such as RSA\cite{RSA}, may be broken by 
quantum computers. 

These new developments enhanced the interest in quantum cryptography
and started a wide surge of interest in the field
of quantum computing.  
However, building such computing devices
is a difficult task, and
quantum computing (which was invented a decade ago ~\cite{Deutsch85})
is only 
doing its first experimental steps.
The building blocks of future quantum computers are one-bit and 
two-bit quantum logical
gates~\cite{Deutsch89,DiV,BDEJ,CZ},
which are currently under intensive 
development~\cite{Haroche,Kimble,Wineland}. 
Building quantum computing 
devices to factor large numbers 
does not seem to be practical in the foreseeable future since it 
requires combining many one-bit and 
two-bit gates. However, 
a single two-bit gate also have intriguing uses in
information processing
and quantum communication such as 
teleporting a quantum state~\cite{BBCJPW}, and dense coding in
quantum cryptography~\cite{BW}. 
We shall show in this paper
that the use of quantum gates together with a quantum
memory
(in which a quantum state can be maintained for a long time without
loss of coherence)
opens new directions in quantum cryptography.
Our system may be practical long before quantum computers are, 
hence provides a short-term application for quantum gates.

One of the main disadvantages of quantum cryptography is its
restriction to relatively short channels. This is due to the fact
that, in contrast to classical channels, a quantum channel cannot use
repeaters to amplify the signal without loss of coherence. 
Currently, working prototypes allow transmission to distances of
about 10 km~\cite{TRT}, 
and up to 23 km for a recent experiment using installed telecom
fibers~\cite{muller}. Commercial systems may become available in the
near future~\cite{LosAl},
so that two users will be able to communicate securely 
(if they are not too far).    
However, building quantum cryptographic {\em networks} based on the
existing 
schemes\footnote{For a suggestion of a quantum cryptographic
networks based
on the existing schemes see~\cite{Townsend}.}
seems to cause severe difficulties (which may even 
make it impractical):    
\begin{enumerate}
\item Quantum communication requires 
any pair of users to
have a common quantum channel, or alternatively a
center (or a telephone-like switching network) connected by 
quantum channels to all the users, which should match any pair of
channels
upon request;
enhancing the security of 
the current world-wide telephone network (which contains about $N
\approx
10^9$ users (telephones) )
using quantum cryptography requires huge investments in quantum
channels and devices.
\item Any user must have the financial and technological abilities
to operate complicated quantum devices.
\item  The keys must be transmitted {\em online}, or else one would
need to transmit $O(N^2)$ keys in advance to enable any pairs of
users to communicate in secrecy.
\item The network must assure authenticity of the users.
\end{enumerate}
It is important to have quantum cryptographic networks  
not suffering from these problems. 

In this work we suggest a new cryptographic scheme 
in which users store quantum states in 
quantum memories, kept in a transmission  center. Upon request from
two users, the center uses two-bit gates to project the product state
of two non correlated particles (one from each user)
onto a fully entangled state. As a result, 
the two users can share a secret
bit, which is unknown {\em even to the center}. 
Our scheme can operate
without quantum channels, if the
quantum states are ``programmed'' at the center. 
In that case, the scheme does not suffer from the four problems just
mentioned, 
and can operate at any distance.
Hence,  
it is especially appropriate
for building a quantum
cryptographic
network of many users.  
Such a system actually shows some of the useful  
properties of the public key cryptosystems,
but yet, it doesn't require computation assumptions.

In Section~\ref{EPR}, we introduce our notation by reviewing various
schemes
for quantum cryptography and specifically
describe the EPR-scheme. We then present  
a new two-party quantum cryptographic scheme
which is a time-reversed EPR-scheme.
In Section~\ref{network}, we present a quantum network based on the
scheme
presented in Section~\ref{EPR} with the addition of the quantum
memories.
In Section~\ref{implementation}, we discuss the possibilities of
implementing our scheme 
in practice.
In Section~\ref{www}, we present a more advanced network, based on
quantum
teleportation, where users can store their states
in different centers and the centers teleport states upon request.
This network uses quantum channels.
However, it requires quantum channels only between the centers,
so that the problems
stated above do not arise.
In Section~\ref{conclusion} we summarize our results.

\section{A Time-reversed EPR-scheme for quantum cryptography}
\label{EPR}
Quantum cryptography provides techniques to distribute keys between
two users, and its safety 
depends only on the fundamental rules of quantum mechanics.
The legitimate users cannot prevent Eve from listening 
to their information exchange, 
but they will know if she does
(hence, in this case, will not use this non-secret information).
The first quantum cryptographic scheme, 
the Bennett Brassard (BB84) scheme~\cite{BB84}, 
was presented already a decade ago.
We describe it using the terminology 
of spin 1/2 
particles, but it can use any two-dimensional Hilbert space.
A classical two-level system, such as a bistable device,
can only be found in one of the two possible states, hence encodes
one bit. In contrast, a quantum system can be prepared in any
coherent superposition of the two basis states, which creates a much
richer structure.  Such a system is now known as a
``qubit''~\cite{ben}
(i.e. quantum bit).
For each qubit, Alice chooses at random whether 
to prepare her state along the $z$ or the $x$ axis, i.e., 
in one of  the two eigenstates of either 
$\hat{S}_z$ or $\hat{S}_x$.
This state, denoted by: $|\! \uparrow \rangle$, $|\! \downarrow
\rangle$, 
$|\! \leftarrow \rangle$ or $|\! \rightarrow \rangle$ is then sent to
Bob. 
It is agreed that the two states 
$|\! \uparrow \rangle$  and $|\! \leftarrow \rangle$ stand for bit
value
`0', 
and the other two states, $|\! \downarrow \rangle$ and $|\!
\rightarrow
\rangle$
 stand for `1'.   
Bob chooses, also at 
random, whether to measure $\hat{S}_z$ or $\hat{S}_x$.  When his 
measurement is along the same axis as Alice's preparation 
(e.g. they both use $\hat{S}_z$), 
the measured value should be the same as hers, whereas when 
they use conjugate axes, there is no correlation between 
his result and Alice's original choice. 
In addition to the quantum channel, the legitimate users also use 
a classical channel which may be monitored, 
but cannot be modified by an eavesdropper (this assumption is
discussed
in \cite{BBBSS}, and is not required if Alice and Bob have a way 
to authenticate each other over the classical channel).
By discussing over this channel Alice and Bob agree
to discard all the instances where they did not use the same
axes. The result should be two strings of perfectly correlated bits.
As the choice of axis 
used by Alice is 
unknown to Eve, any interaction by her will 
unavoidably modify the 
transmission and introduce some errors. 
In practice however, the transmission will never be perfect 
and there will 
be some errors, even in the absence of an eavesdropper. 
Alice and Bob use the classical channel 
to compare some portion of their data and calculate the error rate.
If it is not too high they can 
use classical information processing techniques, such as
error correction and privacy amplification~\cite{BBBSS,BBCM},
to reduce the error rate to zero, while reducing the information
obtained by Eve to zero as well.  
All these operations waste many bits (henceforth, $l$), 
so in order to be left with a key of $L$ bits 
Alice should send  $L' > 2(L + l)$ qubits.
A formal proof of security against an eavesdropper 
who is assumed to be limited only by the rules of quantum mechanics
is still missing 
but may be available soon.
These security aspects are widely discussed in the literature in case
of the
BB84 scheme but are common to all quantum cryptographic schemes 
and will not be discussed here.

More recently, another quantum key distribution scheme, 
based on EPR~\cite{EPR} correlations, 
was suggested by Ekert~\cite{Eke91} and modified by 
Bennett, Brassard and Mermin~\cite{BBM}. 
We describe here the modified version which we call {\em the EPR
scheme}.
In this scheme Alice creates pairs of spin 1/2 particles in the 
singlet  state, 
and sends one particle from each pair to Bob. 
When the two particles are measured
separately the results obtained for them are correlated. 
For example, if they are measured along the same axis, 
the results are opposite, 
regardless of the axis. 
Alice and Bob use the
same sets of axes, say $\hat{S}_z$ and $\hat{S}_x$, and keep
the results only when they used the same axis.   
It is noteworthy that, in the EPR scheme,  the pairs could 
be created by any other party, including Eve herself. 

As this point
will prove crucial in our new scheme, let us discuss it
in more details.  The singlet state may be written in
two ways:
\begin{eqnarray}               
\Psi^{(-)} &=& \sqrt{ \frac{1}{2} }
 \left(|\!\uparrow_A \downarrow_B \rangle -
       |\!\downarrow_A  \uparrow_B
\rangle
\right)  
 \nonumber \\ &=& 
\sqrt{ \frac{1}{2} }
 \left(|\! \leftarrow_A \rightarrow_B \rangle - 
        |\! \rightarrow_A \leftarrow_B \rangle \right)
 \label{singlet-pair} \ ,
\end{eqnarray}           
where the equality follows from 
$|\!  \rightarrow \rangle = \frac{1}{2}(|\!  \uparrow \rangle + |\! 
\downarrow \rangle) \ $ 
and $|\!  \leftarrow \rangle = \frac{1}{2}(|\!  \uparrow \rangle -
|\! 
\downarrow \rangle) \ $,
and where the subscripts `A' and `B', which stand for Alice and for
Bob, 
can be omitted since we always write Alice's particle first.
When Alice and Bob use  the same axis, 
either $z$ or $x$, we use the first or the second equation
respectively,
to see that their measurements always yield opposite results.
The singlet state is the only state which has that property.
Therefore, as Alice and Bob may measure either of these two options, 
any deviation from the protocol by Eve (i.e. any attempt
to create another state), will be detected with non zero 
probability. So Eve must   
create the required singlet state, from
which she cannot extract any information about Alice's
and Bob's measurement (see~\cite{Eke91,BBM} for more details). 

The first aim of our paper is to suggest another scheme for
quantum cryptography, which we shall call {\em the time-reversed
EPR-scheme}.
Let both Alice and Bob send one of the four states of BB84,
$|\! \uparrow \rangle$, $|\! \downarrow \rangle$, 
$|\! \leftarrow \rangle$ or $|\! \rightarrow \rangle$ 
to a third person
whom we refer to as {\em the center} (the purpose of using this name
shall
be clarified in Section~\ref{network}).  
The center measures their qubits together to find 
whether or not the two particles are in a singlet state. 
This can be done
by measuring the total-spin operator $(\hat{S}_{total})^2$. 
If the result of the measurement is $s=0$, then the two particles are
projected onto
the singlet state. In that case Eq.~\ref{singlet-pair}
ensures that,
if the two spins were \underline{prepared} along the same axis,
then they necessarily had opposite values (the projection of the
states with identical spins on the singlet state is zero). 
As result, Bob knows Alice's bit and vice versa. However, from 
Eq.~\ref{singlet-pair}, a honest center, who followed the protocol
and projected onto the singlet state, has absolutely no knowledge on
these bits. For example, when Alice and Bob both used the vertical
axis, the center does not know whether Alice had the up state and Bob
the down state, or vice versa. 
If the measurement result is $s=1$, Alice and Bob cannot infer
anything about the value of each other's bit, and shall discard the
transmission.
The probability of obtaining the singlet state is zero when Alice
and Bob sent the same state (e.g., $\uparrow\uparrow$), 
and is half in case they sent opposite
states. Taking into account the case where Alice and Bob use
different axes (which will also be discarded), we find that the
overall
probability to obtain a usable state is only one eighth.

To create a key with many bits,   
Alice and Bob send strings of quantum states 
($L'>8(L+l)$ qubits) to the center.
The center must be able to keep them for a while (in case the states
do not
arrive at the same time from Alice and Bob), and then to measure the
first pair,
the second pair etc.
The center tells Alice and Bob all cases in which the 
result of the measurement is a singlet,
which happens in one fourth of the cases.
Alice and Bob then compare their axes. 
When they used the same axis
(which happens about half of the time), they know that their
spins are necessarily opposite, and thus Bob can calculate
Alice's bits to share a key with her.
As in the BB84 scheme and the EPR schemes Alice and Bob use the
classical discussion channel to estimate the error-rate.
If it is tolerable they perform error correction 
and privacy amplification, to derive a final $L$-bit key.

The security of our protocol derives from the security of the EPR
protocol, and relies on the fact that the singlet state is the only
state for which the two spins are anticorrelated both in the 
$\hat{S}_z$ and in the $\hat{S}_x$ basis. However, as explained
previously, if the center projects on the singlet state, he does not
get any information on Alice's and Bob's bits. Therefore, a cheating
center needs to project onto a different state (possibly entangled
with his own system), which cannot
give perfect anticorrelations along both  $\hat{S}_z$ and
$\hat{S}_x$ axes. Since the center cannot
know in advance which basis was used by Alice and by Bob (the two
density matrices corresponding to using $\hat{S}_z$ or $\hat{S}_x$
are identical),  he will unavoidably introduce errors, which 
Alice and Bob shall identify  during the discussion. 

In fact, in
terms of eavesdropping possibilities, our protocol and the EPR
protocol are equivalent, as we show using the scheme presented
in Fig.~1.  In this scheme, two
EPR pairs are created, one particle of each pair is sent to the
center, and the
second one to Alice and to Bob. In Fig.~1a, the center performs a
measurement on his two particles first.
A honest center, who follows the agreed protocol, projects the
particles 
onto the singlet
state. The two particles sent to Alice and to Bob are now in the
singlet state as well. This is therefore equivalent to the EPR
scheme. The only difference is that the projection onto the singlet
state performed by the center succeeds with probability 1/4 only.
This means that the center will ask to discard 3/4 of the
transmission, but this does not affect the eavesdropping issue.
A cheating center
can send to Alice and Bob any state he wants, 
including any desired entanglement with his own system, 
by choosing an appropriate unitary transformation 
and the correct state on which to project his own particles. 
To show that, we 
start with the two singlet pairs 
and let the center introduce an ancilla in a state $A_{\rm init}$.
The state of the whole system is:
\begin{equation}  
\Phi_{ABC} =   \frac{1}{2} \Bigl(|\! \uparrow  \downarrow \rangle 
    -  |\!  \downarrow \uparrow \rangle) \otimes
              (|\! \uparrow \downarrow \rangle 
    -   |\! \downarrow \uparrow \rangle \Bigr) \otimes A_{\rm init}
\; .
\end{equation} 
The first particle of each singlet pair is sent to Alice and to Bob
respectively, while the center keeps the second, together with his
ancilla. The state $\Phi_{ABC}$ can thus be rearranged as:
\begin{equation}
\Phi_{ABC} = 
 \frac{1}{2} \Bigl( |\! \uparrow \uparrow \rangle_{AB} \otimes
              |\! \downarrow \downarrow \rangle_{C} 
       +      |\! \downarrow \downarrow \rangle_{AB} \otimes
              |\! \uparrow \uparrow \rangle_{C} 
       -      |\! \uparrow \downarrow \rangle_{AB} \otimes
              |\! \downarrow \uparrow \rangle_{C}
       -      |\! \downarrow \uparrow \rangle_{AB} \otimes
              |\! \uparrow \downarrow \rangle_{C}           \Bigr)
              \otimes A_{\rm init} \; , 
\end{equation}
where the index $AB$ refers to the particles sent to Alice and Bob,
and the index $C$ refers to the particles kept by the center. 
The center now applies a unitary transformation $U$ to entangle his
particles with the ancilla in the following way:
\begin{eqnarray} 
U \Phi_{ABC}  =  \frac{1}{2} \Bigl( & &
               |\! \uparrow \uparrow \rangle_{AB} \otimes
              |\! \downarrow \downarrow \rangle_C  \otimes A_1
       +      |\! \downarrow \downarrow \rangle_{AB} \otimes
              |\! \uparrow \uparrow \rangle_C \otimes A_2 
\nonumber \\               
     &   & 
      -        |\! \uparrow \downarrow \rangle_{AB} \otimes
              |\! \downarrow \uparrow \rangle_C  \otimes A_3
       -      |\! \downarrow \uparrow \rangle_{AB} \otimes
              |\! \uparrow \downarrow  \rangle_C \otimes  A_4 \Bigr)
\; ,
\end{eqnarray}
with $A_i$ any normalized states of the ancilla which are (in general)  
not orthogonal to one another.
By projecting his state onto 
$ \psi =   
   \alpha |\! \downarrow \downarrow \rangle_C +
   \beta  |\! \uparrow \uparrow \rangle_C - 
   \gamma  |\! \downarrow \uparrow \rangle_C - 
   \delta  |\! \uparrow \downarrow \rangle_C \ \  $
(this projection succeeds with probability 1/4 on average), the center creates
the state
\begin{equation}  \Psi_{ABC} =  \frac{1}{2} \Bigl(
          \alpha^*  |\! \uparrow \uparrow \rangle_{AB} \otimes A_1
        + \beta^*   |\! \downarrow \downarrow \rangle_{AB} \otimes A_2
        + \gamma^*  |\! \uparrow \downarrow \rangle_{AB} \otimes A_3 
        + \delta^*  |\! \downarrow \uparrow \rangle_{AB} \otimes A_4  
\Bigr) \; , 
\end{equation}
which is the most general state the center could create when cheating
the EPR scheme~\cite{BBM}. This demonstrate the equivalence between
Fig.~1a and the EPR scheme.

In Fig.~1b, the first measurement is performed by Alice and
Bob, who project the particles onto the BB84 states. Therefore, the
particles arriving at the center are also in the BB84 states, and
this scheme is identical to ours. Since the relative time of the
measurements cannot influence the outcome, all these schemes are
equivalent.
Following the same reasoning, but in two steps (first letting only
Alice measure before the center) it is also possible to show that 
the security of the BB84 scheme implies the security of our scheme.
Since the security of the EPR scheme implies
the security of the BB84 scheme~\cite{BBM}, our proof actually 
shows that 
the security of the three
schemes is equivalent.

Using only the total spin measurement, less than one eighth of the 
qubits could be used. 
A better choice, although 
possibly more difficult to implement in practice, 
is to measure the Bell operator
(defined in~\cite{BMR}) whose eigenstates (the {\em Bell states})
are the singlet state, 
 $\Psi^{(-)}$ (equation \ref{singlet-pair}),
and the three other states:
\begin{equation}    
 \Phi^{(+)} 
 = \sqrt{\frac{1}{2}}
 \left(|\!  \uparrow \uparrow \rangle +  
|\!\downarrow \downarrow \rangle
\right)  
= \sqrt{ \frac{1}{2} }
 \left(|\! \leftarrow \leftarrow \rangle + 
|\! \rightarrow \rightarrow \rangle \right) 
\; , \label{phiplus} \end{equation}   
\begin{equation} 
\Psi^{(+)} = \sqrt{ \frac{1}{2} }
 \left(|\!  \uparrow \downarrow \rangle +  
|\! \downarrow \uparrow \rangle \right) 
= - \sqrt{ \frac{1}{2} }
 \left(|\! \leftarrow \leftarrow \rangle - 
|\! \rightarrow \rightarrow \rangle \right) 
   \label{psiplus} \end{equation} 
and
\begin{equation}
\label{phiminus}
 \Phi^{(-)} = \sqrt{\frac{1}{2}}
 \left(|\!  \uparrow \uparrow \rangle -  
\downarrow \downarrow \rangle \right)  
= \sqrt{ \frac{1}{2} }
 \left(|\! \leftarrow \rightarrow \rangle + 
|\! \rightarrow \leftarrow \rangle \right)  \ ,
\end{equation}   
where the second expression for each of the Bell states is derived by
expanding them (as was done for the singlet state)
into the 
$(|\! \leftarrow \rangle \, , \, |\! \rightarrow \rangle )$ basis.
Consider a case where Alice and Bob used the same basis.
According to the result of the measurement, and to the
choice of axes by Alice and Bob, their prepared states are known to
be either
correlated (e.g. if the result is $\Phi^{(-)}$ and they both used the
$z$ axis), or anticorrelated (e.g. if the result is still
$\Phi^{(-)}$ but they both used the $x$ axis).

The protocol goes as
follows:
\begin{itemize}
\item
The center retrieves the particles from Alice and Bob and 
measures the Bell operator on each pair. He gets 
one of the 
four above states, and tells his result to Alice and Bob. 
\item  Alice and Bob 
tell each other the axis they used (but not the bit value). When they
used different axes, they discard the transmission. Whenever
they used 
the same axis, they know if their bits are correlated 
or anticorrelated. In this case
half of the quantum states are used 
to derive the desired key, and $L'>2(L+l)$ qubits are required. 
\end{itemize}

The proof of security for this case is similar to the proof 
in the singlet case. A honest center, who projects the states onto
the allowed states, cannot get any information on the bits. For
example, if the center obtains the state $\Phi^{(+)}$, and Alice and
Bob 
announce later that they used the horizontal axis, the center only
knows that
either both Alice and Bob have the left state, or both have the right
state. But he cannot know which of these two possibilities occurred,
hence has no information on the bit values. Moreover, similarly to
the singlet case, $\Phi^{(+)}$ is the only state for which Alice's
and Bob's states have such correlations along both $x$ and $z$ axes.
Therefore, a cheating center, who needs to 
create  a different state in order to gain information, shall be
detected with 
finite probability. 

\section{A quantum cryptographic network}
\label{network}
In this Section we combine the reversed EPR scheme  
and the use of quantum memories into a classical network to
present a {\em quantum cryptographic network}.
The classical protocol for a network uses a ``hidden file''
managed by a communication center. 
Any user is allowed to put data (secret keys) in the file, under his
name, but 
only the center has access to the data. 
Let there be $N$ users, and let each of them store many $L$-bits
strings.
Upon request from two users, the center uses their data and creates 
a secret key for them, which is 
shared by both of them:
the center calculates the XOR of one 
string of the first user (say, $a_1 \ldots a_L$ of Alice),
and one string of the second user (say, $b_1 \ldots b_L$ of Bob); 
the XOR of a string is calculated bit by bit using  
$c_j = a_j \oplus b_j$
(the parity of the two bits),
and the resultant string, $C = c_1 \ldots c_L$, is transmitted 
(via a classical  unprotected channel) to Alice;
Alice rederives Bob's string by calculating the XOR of  
her string with the received 
string, and can use Bob's string as their common key.
Secure transmissions from each user to the center can be done either
by personal delivery, trusted couriers
or quantum key distribution.
Such a classical key distribution scheme  
is perfectly secure if we 
assume that the center holding them  
is perfectly safe and trusted. 
No other person (except 
the center) can have any information on their key.
Even a powerful eavesdropper who can impersonate the center and all
the  
users cannot eavesdrop, since the center and each of 
the legitimate users can use 
some of the secret bits for authenticating each other.
Alice and Bob need to trust the center for two different purposes:
\begin{enumerate}
\item To ``forget'' their secret key (and not trying to listen to the
messages
transmitted using that key);
\item To authenticate one to the other in case they 
have no other way of authentication. 
This is a new possibility of authentication, added to the two
options, previously 
mentioned in Section~\ref{EPR}.
Thus the assumption
of having classical channels which cannot be modified 
can be completely removed, 
even if the users have no other way to authenticate each other.
\end{enumerate}
The main reason why this simple scheme is
not satisfactory in practice, is that it concentrates too much power
in the distribution
center. Indeed the center can understand all the secret
communications going
through 
its distribution web, or connect Alice to an adversary instead of to
Bob. 
Even if we assume that the center 
is trusted,
any eavesdropper who manages to get access to it could decipher all
the 
communications.  

Using a quantum memory instead of a classical memory is the key-point
in deriving the quantum network, hence we present it in more details.
While a classical bit can only represent a zero or a one, a qubit
$| \phi\rangle = \alpha | 0\rangle + \beta | 1\rangle$ is
described
by two complex numbers (up to freedom of overall phase and 
normalization requirement).
Unlike a classical memory which keeps $n$ classical independent bits,
the $n$ qubits in a quantum memory can have non-classical
correlations,
and the state of a quantum register is described by $2^n$ complex
numbers
(up to freedom of overall phase and normalization requirement).
Even the simplest form of a memory (where the qubits are never
correlated)
is very important in quantum cryptography. 
For example, it allows 
doubling the efficiency of the BB84 scheme to use only $L' > L+l$
qubits: instead of measuring the state sent by Alice immediately,
Bob keeps it in a quantum memory; waits for Alice to disclose
her basis; and then measures the state in the correct basis.
In this case, 
the BB84 scheme can be used
directly to transmit messages instead of random keys; Alice decide in
advance
which qubits will be used for error estimation, and encode the
message
using the rest of the qubits and using 
block-coding techniques to allow for error-correction.
A quantum memory is also a basic tool for eavesdropping attacks,
as it allows Eve to couple the transmitted states to an ancilla, and
delay the measurement on the ancilla till the public exchange of
basis.

We now present a quantum key 
distribution network which uses 
a {\em quantum file} instead of the classical hidden file,
and {\em removes} the
requirement of a trusted center. 
Alternatively, we can release the usual
assumption of quantum cryptography \cite{BBBSS}
 -- that classical channels cannot be 
modified by Eve -- if we are willing 
to trust the center
for authentication (without trusting him for "forgetting" their
qubits).
Instead of storing $L$ classical bits to make a future
key, each user shall store $L'$ quantum states (qubits) in specially
devised
quantum memories kept in a center. 
Upon request from two users, the center 
performs the time reversed EPR scheme described in the previous
Section and creates correlations between the bits. 
The resulting
string C, which holds the correlation data is sent 
to Alice via a classical channel.
As in the classical case, using string C, Alice can calculate Bob's
string
to derive a final common key of $L$ bits.
If Alice and Bob compare bases {\em after} deriving the data from the
center, 
then, as explained in Section~\ref{EPR}, any attempt by the
center to obtain the value of these bits will create errors and be
discovered by Alice and Bob. The center therefore does not need to be
trusted anymore. 
Unlike other quantum schemes, the actual (online) distribution of the
secret
keys is performed on classical channels. First, the center let Alice
and Bob
know the state he got. Then, Alice and Bob continue as in the other
two schemes
previously described to obtain the final key.
All the quantum communication is done in advance,
when the users ``deposit'' their quantum strings in the center
(preferably in a personal meeting).

When $L$-bit strings are stored in a classical hidden file, 
two users derive $L$-bit strings of correlated bits. 
Using quantum states for representing the bits, longer strings of
length $L'>L+l$ are required, 
since some bits will be used for error estimation,
error correction and 
privacy amplification.
The exact ratio depends on the expected error-rate in the channel.
Only the bits which are encoded in the same basis by both users 
can be used, therefore $L' > 2(L+l)$ bits are actually required 
(we assume that the more 
efficient scheme of measuring the Bell basis is used).

Let us summarize the protocol as follows:
\begin{itemize}
\item In the preparation step the user sends (gives) 
\mbox{$L'$-bit} strings
to the center, each bit is represented by  
one of the four states of the BB84 protocol.
The center keeps these quantum states in a quantum file without
measuring them. It is important that the system used for keeping the
quantum states will preserve them for a long time (as long as
required
until the actual key distribution is performed). 
\item 
When Alice and Bob wish to obtain a common secret key, they ask the
center to 
create correlations between two strings, one of Alice and one of Bob.
The center performs the Bell operator measurement 
on each pair of photons, which  projects them onto one of the Bell
states, and tells Alice and Bob the result he obtained. 
After Alice gets the results from 
the center (and not before that),  
Alice and Bob compare the basis they used and keep only 
the bits for which they used the same basis. 
In this case, and according to the state obtained, the  states of
Alice and Bob are either correlated or anticorrelated. So, Alice for
example inverts all her bits which should be anticorrelated with
Bob's. The remaining string should be identical with Bob's, apart
from possible errors.
\item A honest center, who performed the correct projections on the
Bell states does not get any information on the string.
\item A cheating center (or any other eavesdropper who might have had
access to the quantum files), who modified the allowed states,
unavoidably introduced errors between the two strings.
\item Alice and Bob perform error estimation, error correction
and privacy amplification to derive a final key.
\end{itemize}

The quantum channel 
is used only as a preparation step between each user and the center, 
and all the online communication 
is done via a classical channel.
Yet, only $O(N)$ keys 
are required 
to enable secret communication between 
any pair of the $N$ users.
Any other quantum key distribution scheme requires 
$O(N^2)$ keys, or else requires online quantum communication.
In fact, our scheme does not require quantum {\em channels} at all.
As in old implementations of quantum cryptography~\cite{BBBW}, 
the four quantum states can be chosen in any 2-dimensional 
Hilbert space.
Instead of sending them, each user could arrive once in a while to 
the center, and ``program'' his states into the quantum file.
If the memory can keep the states unperturbed long enough then each
user
can put as many strings as he needs till his next visit to the
center.
By using personal delivery of the quantum states we
replace the distance
limitation of all other schemes by a time limit, 
and solve the problems
of a quantum cryptographic network which were described in the
introduction.
All the technically
involved steps, such as storing qubits and 
performing Bell measurements
occur only at the center.

\section{Implementations}
\label{implementation}
Our scheme requires the possibility to program, 
store and manipulate quantum bits rather than to transmit them.
Therefore any 
2-dimensional Hilbert space system can be considered and this 
opens a variety of possible implementations.
Fortunately, almost the same requirements appeared recently in
quantum computing, and are being thoroughly
investigated
by both theorists 
\cite{Deutsch89,DiV,BDEJ,CZ},
and experimentalists
\cite{Haroche,Kimble,Wineland}. 
The main difference in the requirements is that the quantum bits in
our
scheme are subjected only once to a unitary operation of calculation
hence the problem
of decoherence is much less severe.

We estimate\footnote{The following suggestions were investigated
with the help of D. DiVincenzo~\cite{DD}.} that 
it may be possible to implement a working prototype of 
our scheme within a few years, 
with small modifications of existing technology.
Such a prototype shall be able to keep quantum states for
a few minutes 
and to allow to perform two-bit operations on them. 
At the moment, the best candidates for combining these two operations 
are ion traps.
In ion traps the quantum bits can be kept in internal degrees of
freedom 
(say, spin) of the ions, and
in phononic degrees of freedom of few ions together.
It is already possible to keep quantum states in the spin 
of the ions  
for more than 10 minutes~\cite{Wineland2} 
and in principle it is possible 
to keep them for years.
These ion traps are also the among the best candidates for two-bit
operations 
since there are ways to use the phononic degrees of freedom to
perform two-bit operations~\cite{CZ}.
Barenco et al.~\cite{BDEJ} realized that
a single quantum controlled-NOT logical gate would be 
sufficient to perform the Bell measurement. 
Using ion traps it is possible to (partially) perform the Bell
measurement 
as shown both in theory~\cite{CZ} and in an experiment
\cite{Wineland}.
Combining together the two experiments to have
both long lived quantum states and the possibility to manipulate them
will allow two users to derive a few  
secure common bits.

The way to establish a real working scheme with a center and 
with many users is still long.
The main obstacle is that it is currently impossible to transfer
a quantum state from one ion trap to another.
A real network should allow each user to program his quantum states
in a register (say, at least one separated ion trap for each user).
Then, upon request of two users, the center should be able to move
one bit of each of the two users to another ion trap where he can
perform
the Bell measurement without disturbing the other quantum states.
Recently, the possibility of doing this arose from the
idea of combining
ion-traps (where the ions are well controlled) and QED-cavity
together,
and to use the same internal degrees of freedom for 
both~\cite{DD,Pellizz}. 
We shall call this combination {\em cavitrap} for convenience.
In QED cavity \cite{Haroche} the internal degrees 
of freedom are coupled to photons and not to phonons.
Recently, another group \cite{Kimble}
has shown that it is possible to use polarization
states of
photons instead of using the $| 0\rangle$ and $| 1\rangle$ Fock
state.
If such photon states are used in a cavitrap, it may be possible to
use 
them to transmit a quantum state from
one cavitrap to another~\cite{DD}.
In some sense, this will be an implementation of the nuclear spins
based
``quantum gearbox'' suggested by DiVincenzo~\cite{DiV}.

All this discussion would be nothing but a fantasy if our scheme
would
yield the only application of these ideas.
However, as we already said, similar (and more complicated) ideas 
are required for other usages of
quantum gates
in both quantum computing and quantum information,
and a lot of effort is invested in both the theory and application 
of quantum gates.  
Quantum memory is less discussed in quantum computing,
but more in quantum cryptography,
and the use of quantum memory to attack quantum cryptographic schemes
is already appreciated. 
Nevertheless, the use of a quantum memory 
by the legitimate users of
a quantum cryptographic scheme is not common, due to the desire to
present protocols which may be implemented with existing technology.
However, bypassing this ``habit'' and using a forthcoming tool is
certainly 
justified when it allows to carry out new and important
tasks which cannot
be performed without it. 
The use of quantum memory in our network is clearly such a case,
and we are sure that it will allow many
other new tasks in the future.
While we were working on this paper, it was suggested to 
use both quantum gates and quantum memory 
for purification of singlets~\cite{B-et-al}, an idea which
is useful for quantum cryptography as well.

\section{World-wide network of many centers}
\label{www}

The network of Section~\ref{network} is well suited for 
communication
among users who
are not far away and can arrive to the center.
For two users who are far away and cannot come to the same center
our network may not be appropriate.
We now show that it can be modified to be useful also in this case.
Let there be many centers, and many users in each center.
Each two centers should share many EPR singlet pairs
and upon request of users of the two centers, one center would
teleport \cite{BBCJPW} the
qubits of his user to the other center. This operation can be done
with 100\% efficiency (but it may, however, increase the error-rate).
The singlet pairs can be transmitted using any quantum cryptography 
scheme or even using a teleportation scheme and
a {\em supercenter} who share EPR pairs with all centers. 
However, the transmission and distribution of singlet pairs require
quantum channels even if done in advance\footnote{And transporting 
quantum states to deliver them by a personal meeting is similar
to transmitting them through channels}.

Do we lose all the benefits we gained before?
Certainly not.
Still, only the centers need to have ability of performing quantum
operations.
All quantum transmission is done in advance, and yet,
there is no need for $O(N^2)$ strings, since the number of centers
is much smaller than the number of users.
Authentication is still simple.
One problem of any quantum channel is the limit on its length.
Our first scheme (with one center) replaced it by  a limit on time. 
It would be bad to have both problems in a network of many centers.
However, the suggestion of purifying singlets~\cite{B-et-al}
enables transmitting signals to longer distances.
Our scheme can make an excellent use of it, since only the
centers
need to have the technological ability of purifying singlets.
Moreover, several transmission stations can be put in between to
improve 
transmission (this idea was suggested by DiVincenzo~\cite{DD}) by
performing purification of singlets between any two neighboring
stations
and then use teleportation from one station to the next to derive
purified singlets shared by the centers.
   
\section{Summary and Discussion}
\label{conclusion}

In this paper we introduced a new scheme for quantum cryptography
based on a time-reversed EPR scheme.
We suggested two new types of networks:
a classical one based on hidden files, and one based on quantum
files.
The security of key distribution protocols in these networks 
does not rely on computational complexity assumptions.
Both networks can be used 
to distribute keys in a secure way among any two users using 
simple online communication via classical channel.
In the case of hidden files it is done with the help
of a trusted center who can have access to all the information
exchanged through its lines. 
Using quantum files, the center need not be trusted.
Users have the means to
check whether the center, or any other eavesdropper, tried to obtain
information on the transmitted messages. 

The one-center quantum network we suggest
does not require any quantum channel 
at all, and can be implemented in a center where 
each user ``programs'' 
his states into a quantum memory.
We estimate that a working prototype may be built in the near future
using ion-traps technology.
A real network can be built when the problem of transmitting 
a quantum state from one trap to another is be solved (perhaps
using
cavitrap with polarization states).
The machinery required for our scheme is also required for 
much more complicated tasks such as purification and quantum
computing.
In this respect, our scheme may represent a first practical 
application for these new devices, which are now being planned 
in various laboratories. 
We hope that our work will motivate more research for systems which
can both
keep a quantum state for a long time, and allow for the desired
programming
and measurements.
We didn't pay much attention to the delicate problem of programming
the states.
The programming requires a simple equipment to make sure that the
center does not
eavesdrop on the preparation step. While cavitraps are very
complicated,
programming the polarization
state of each of the photons may be quite simple in the future.

A future system of secure communication based on the protocol
of Section~\ref{www}
would involve a number of large transmission 
centers, which can exchange EPR correlated particles and 
store them, together with the qubits deposited at the center
by various users. Secure communication between any pair
of users would then imply teleportation of the states to 
the same center, followed by the creation of correlations
between the two strings. 

\acknowledgments 
We thank A. Peres for many discussions and we also thank
C. H. Bennett, G. Brassard, C. Cr\'epeau, A. Ekert and 
L. Vaidman for helpful remarks.
Especially we would like to thank D. DiVincenzo for his great help in
analyzing the possible implementations of our protocol.
B.H. gratefully acknowledges financial support from the Swiss Fonds
National de Recherche Scientifique.

\begin{figure}[h]
\caption{Two processes, which we use to prove 
the security of the protocol. 
In both figures, one 
particle of each EPR correlated pair (denoted by dashed lines)
is sent to the center, 
who performs a Bell measurement. We consider only the case
where the result of the measurement is a singlet state.
The second
particles are sent to Alice and to Bob respectively, 
who project them 
onto the BB84 states. 
In a), the first measurement is done by
the center. The particles arriving to Alice and Bob 
are therefore in the singlet state like in the EPR based 
protocol. In b), the first measurement is performed by 
Alice and Bob. Each particle sent to the center is 
therefore in one of the BB84 states. This is similar to
our protocol.}
\end{figure}
\end{document}